\documentclass[11pt]{article}
\usepackage{moriond,epsfig}
\usepackage{amsbsy}
\usepackage{amssymb}
\usepackage{amstext}

\def\be{\begin{equation}}
\def\ee{\end{equation}}
\def\bea{\begin{eqnarray}}
\def\eea{\end{eqnarray}}

\newcommand{\ve}{\varepsilon}

\begin{document}
\vspace*{4cm}
\title{DATA ANALYSIS FOR CONTINUOUS GRAVITATIONAL-WAVE SIGNALS}

\author{ A. KR\'OLAK }

\address{Institute of Mathematics, Polish Academy of Sciences,
\'Sniadeckich 8,\\00-950 Warsaw, Poland\\and\\
Astronomical Center, Pedagogical University, Lubuska 2,\\
Zielona G\'ora, Poland}
~
\maketitle\abstracts{
The main problem that we will face in the data analysis for
continuous gravitational-wave sources is processing of
a very long time series and a very large parameter space.
We present a number of analytic and numerical tools that can be useful
in such a data analysis. These consist of methods to calculate
false alarm probabilities, use of probabilistic algorithms, application of
signal splitting, and accurate estimation of parameters by means of
optimization algorithms.}

\section{Introduction}\label{sec:intro}
The interest in the data analysis for continuous
gravitational-wave signals is growing. A prime example of a source of such
signals is a spinning neutron star. These may be the first
signals to look for in data streams from the long-arm laser
interferometers and they are attractive sources for currently
operating bar detectors. A number of theoretical papers were published
on various data analysis schemes and a certain amount of real data
were analyzed involving data from the prototypes of laser interferometers
and from the bar detectors. Limited space precludes a review of available
literature on this subject. The theoretical research clearly shows
that data analysis will involve processing of very long time series
and exploration of a very large parameter space. It turns out that
optimal processing of month long data streams that we need to ensure
a reasonable chance of detection of such signals is
computationally prohibitive~\cite{BCCS98} and methods to reduce the computational
burden must be found.
In this contribution we present a few data analysis tools that can
be useful in processing data for continuous signals and that were
a subject of our recent research.
We do not offer a complete end-to-end data analysis scheme.

\section{Data analysis tools for continuous signals}\label{sec:tools}
\subsection{Significance of detected events}\label{subsec:sig}
We assume that the noise $n$ in the detector is an additive, stationary,
Gaussian, and zero-mean continuous in time $t$ random process.
Then the data $x$ (if the signal $h$ is present) can be written as
\be
x = n + h.
\ee
We have shown~\cite{P2,P3} that the gravitational-wave signal from
a spinning neutron star can be approximated by the following signal
\be
\label{cal1}
h(t;h_o,\Phi_0,\boldsymbol{\xi})
= h_o \sin\left[\Phi(t;\boldsymbol{\xi})+\Phi_0\right],
\ee
where
\bea
\label{cal2}
\Phi(t;\boldsymbol{\xi}) &=&
2\pi\sum_{k=0}^{s} f_k \left(\frac{t}{T_o}\right)^{k+1}
\nonumber\\&&
+\frac{2\pi}{c}\left\{\alpha_1
\left[R_{ES}\sin\left(\phi_o+\Omega_o t\right)
+R_{E}\cos\lambda\cos\ve\sin\left(\phi_r+\Omega_r t\right)\right]
\right.\nonumber\\&&\left.
+\alpha_2
\left[R_{ES}\cos\left(\phi_o+\Omega_o t\right)
+R_{E}\cos\lambda\cos\left(\phi_r+\Omega_r t\right)\right]\right\},
\eea
and where $T_o$ is the observation time, $R_{ES} = 1$AU is the mean
distance from the Earth's center to the SSB, $R_E$ is the mean
radius of the Earth, $\Omega_o$ is the mean orbital angular
velocity of the Earth, $\phi_o$ is a deterministic phase which
defines the position of Earth in its orbital motion at $t = 0$, $\epsilon$
is the angle between the ecliptic and the Earth's equator, $\lambda$
is detector's latitude.
The vector $\boldsymbol{\xi}=(\alpha_1,\alpha_2,f_0,\ldots,f_s)$,
so the phase $\Phi$ depends on $s+3$ parameters,
$h_o$ and $\Phi_o$ are respectively constant amplitude
and phase. The parameters $\alpha_1$ and $\alpha_2$
are defined by
\be
\label{a1a2}
\alpha_1 := f_0
\left(\cos\ve\sin\alpha\cos\delta+\sin\ve\sin\delta\right),\quad
\alpha_2 := f_0 \cos\alpha\cos\delta,
\ee
where $\alpha$ is right ascention and $\delta$ is declination of the source.

From the maximum likelihood principle we find that in order to detect the
signal we need to compute the following statistics
\be
\label{Sstat}
{\mathcal F}(\boldsymbol{\xi}) = \frac{2T_o}{S_h(f_0)}
\left| F \right|^2,
\ee
where
\be
F = \int^{T_o/2}_{-T_o/2} x(t)
\exp\left\{-i \Phi(t;\boldsymbol{\xi})\right\}\,dt,
\ee
and where $S_h(f_0)$ is the one-sided spectral density at frequency
$f_0$. We assume that we can neglect variations of $S_h$ over the
bandwidth of the signal.
The statistics ${\mathcal F}(\boldsymbol{\xi})$ is a certain generalized
multiparameter random process called the {\em random field}~\cite{A81}.
If the vector
$\boldsymbol{\xi}$ is one-dimensional the random field is simply a random
process.
When the signal is absent the above random
field ${\mathcal F}$ is {\em homogeneous} i.e. its mean
$m$ is constant and the autocovariance function $C$ depends only on the
difference $\boldsymbol{\tau} := \boldsymbol{\xi}-\boldsymbol{\xi}'$.
\bea
\label{covdef}
C(\boldsymbol{\xi},\boldsymbol{\xi}') &:=&
E\left\{[{\mathcal F}(\boldsymbol{\xi})-m(\boldsymbol{\xi})]
[{\mathcal F}(\boldsymbol{\xi}')-m(\boldsymbol{\xi}')]\right\} =
C(\boldsymbol{\tau}).
\eea
The homogeneity of the random field is a direct consequence
of the linearity of the phase of the signal in parameters.
Moreover one can show that for the case of a Gaussian noise the
field ${\mathcal F}$ is a (generalized) $\chi^2$ field~\cite{P3}.
To assess the significance of events we need to calculate the
probability the statistics ${\mathcal F}$ crosses a certain threshold
when the data is only noise~\cite{P3} which is called the {\em false
alarm probability}. We have developed two approaches to calculate
this probability: one based on the concept of an elementary cell
of the parameter space and the other on the geometry of the
random fields.
\subsubsection{Cells}\label{subsubsec:cell}
The correlation function $C(\tau)$ has a maximum at $\tau = 0$
and we can define the characteristic correlation hypersurface by
the equation
\be
\label{gcov1}
C(\boldsymbol{\tau}) = \frac{1}{2}C(0).
\ee
The volume of this hypersurface defines an {\em elementary cell}
in the parameter space and we can calculate the number of cells
$N_c$ from the formula
\be
\label{NT}
N_c = \frac{V^{\text{total}}}{V^{\text{cell}}}.
\ee
where $V_{\text{total}}$ is the volume of the parameter space.
We approximate the probability distribution ${\mathcal F}(\boldsymbol{\xi})$ in
each cell by probability $p_0({\mathcal F})$ when the signal parameters are known
(in our case $p_0({\mathcal F}) = \exp(-{\mathcal F})$).  The values of the
statistics ${\mathcal F}$ in each cell can be considered as independent random
variables.  The probability that ${\mathcal F}$ does not exceed the threshold
${\mathcal F}_o$ in a given cell is $1 - \exp(-{\mathcal F}_o)$.
Consequently the probability
that ${\mathcal F}$ does not exceed the threshold ${\mathcal F}_o$ in all the
$N_c$ cells is $[1 - \exp(-{\mathcal F}_o)]^{N_c}$.  The probability $P^c_F$ that
${\mathcal F}$ exceeds ${\mathcal F}_o$ in {\em one or more} cell is given by
\be
P^c_F({\mathcal F}_o) = 1 - [1 - \exp(-{\mathcal F}_o)]^{N_c}.
\label{FP}
\ee
This is the false alarm probability when the phase parameters are unknown.
\subsubsection{Geometry of random fields}\label{subsubsec:geom}
The second approach to calculate the false alarm probability involves
calculation of the number of threshold crossings by the
statistics ${\mathcal F}$.
We need to count somehow the number of times a random field crosses a fixed
hypersurface.  Let ${\mathcal F(\boldsymbol{\xi})}$ be $M$-dimensional
homogeneous real-valued random field where parameters $\boldsymbol{\xi} =
(\xi_1,\dots,\xi_M)$ belong to $M$-dimensional Euclidean space ${\mathbb R}^M$
and
let $\boldsymbol{K}$ be a compact subset of ${\mathbb R}^M$.  We define the
{\em
excursion set} of ${\mathcal F(\boldsymbol{\xi})}$ inside $\boldsymbol{K}$
above the level ${\mathcal F}_o$ as
\be
A_{\mathcal F}({\mathcal F}_o,\boldsymbol{K}) := \left\{
\boldsymbol{\xi}\in\boldsymbol{K}: {\mathcal F}\geq{\mathcal F}_o
\right\}.
\ee
It was found~\cite{A81} that when the excursion set does not intersect the
boundary of the set $\boldsymbol{K}$ then a suitable analogue of the mean
number of level crossings is the expectation value of the Euler characteristic
$\chi$
of the set $A_{\mathcal F}$.  For simplicity we shall denote
$\chi[A_{\mathcal F}({\mathcal F}_o,\boldsymbol{K})]$ by $\chi_{{\mathcal
F}_o}$.
It turns out that using the Morse theory the
expectation values of the Euler characteristic of $A_{\mathcal F}$ can be
given in terms of certain multidimensional integrals~\cite{A81}
and close form formulae could be obtained for homogeneous
$M$-dimensional Gaussian fields and 2-dimensional $\chi^2$ fields~\cite{A81}.
Recently Worsley~\cite{W94}
obtained explicit formulae for $M$-dimensional homogeneous $\chi^2$ field.
One finds that asymptotically (i.e. for large thresholds)
the false alarm probability can be approximated by
\be
\label{FPg}
P^g_F \cong \frac{({\mathcal F}_o/2)^{M/2}}{\Gamma(M/2+1)} P^c_F,
\ee
where $P^c_F$ is given by Eq.~\ref{FP}.
Our numerical simulations have shown that the formula for the false alarm
probability based on the cell concept (Eq.~\ref{FP})
applies when the statistics
${\mathcal F}$ is coarsely sampled by means of FFT whereas formula
based on the expectation of the Euler characteristic (Eq.~\ref{FPg})
applies when the statistics is finely sampled.
Using Eq.~\ref{FP} We have found that for observation time $T_o$
of $7$ days and extremal values of spindowns estimated form
$|{f_k}|_{max} = k!|\frac{f_{max}}{\tau^k_{min}}|$ where maximum
frequency $f_{max}$ is 1kHz and minimum spindown age $\tau_{min}$
is taken as $10^3$yr the number of cells is $10^{16}$ what
corresponds to $10$ Teraflop computing power when we want to keep
the data processing rate equal to data acquisition rate.
In numerical calculation we expand the left hand side of
Eq.~\ref{gcov1} in Taylor series keeping only terms to the 2nd
order. This gives approximation of the hypersurface of correlation
by a hyperellipse.

It is interesting to note that the above approach has been used in
the search for significant signals in the cosmic microwave
background maps~\cite{T94} and brain tomography images~\cite{W94}.

\subsection{Probabilistic algorithms}\label{subsec:prob}
Our estimate at the end of the previous section shows that the number of
independent cells to search for continuous signals is very large
and computing power required huge. However in data analysis for
gravitational-wave signals we would first like to know
whether any gravitational-waves are present. Our primary objective
is to detect the first gravitational-wave signal. To do so
one does not need to search the whole parameter space.
A related problem occurs in testing whether a certain very large
number is a prime. Primes are very important in designing
encryption schemes. Here a brilliant solution in the form of
Miller-Rabin test~\cite{R76} was found with the
help of probabilistic algorithms. We want to test that a certain very large
natural number $n$ is a prime. We choose randomly a natural number $b_i$
less than $n$. We then perform the Miller
test which shows for some numbers $b_i$ with certainty that
$n$ is not a prime. If test is passed $b_i$ is called witness of
nonprimality. Then by lemma due to Rabin we know that if $n$ is
not a prime at least 1/2 of the positive integers between 1 and n
are witnesses. Thus probability $P_D$that $n$ is a prime after $m$
random choices of $b_i$ is $\geq 1 - 1/2^m$ and thus for very
modest values of $m$ is very high. Of course we only know that $n$
is a prime with a certain (very high) probability. However this is good
enough for many applications. In the case of signal detection in
noise we can only detect the signal with the certain probability
thus use of probabilistic algorithms is not a limitation.
However here we are only able to offer a very naive application
of probabilistic algorithms: choose the values of parameters of
the templates randomly from the parameter space  instead of
searching systematically the parameter space. If we have $N_c$
cells in the parameter space and there are signals in $N_E$ cells the
probability of detection of the signal after $m$ random choices of
a cell to search is given by
\be
P_D = 1 - (1 - \frac{N_E}{N_c})^m.
\ee
Assuming that $\frac{N_E}{N_c} << 1$ we find that we need
$\frac{N_c}{N_E}$ random choices of cells in order that detection probability
is $>90\%$. This can lead to a substantial reduction of the
computational cost to detect the first signal with respect to
the cost to search the whole parameters space. For example for the
case given at the end of the previous section if the number of
signals is $10^3$ the computational power decreases to a manageable amount
of 10 Gflops.
\subsection{Signal splitting}\label{subsec:split}
In this section we would like to discuss a technique that can distribute
the data processing into several smaller computers.
We shall call this technique {\em signal splitting}.
We can divide the available bandwidth of the detector
$(f_{\text{min}},f_{\text{max}})$ into $M$ adjacent intervals of length $B$.
We then apply a standard technique of heterodyning.  For each of the chosen bands
we multiply our data time series by $\exp(-2\pi i f_I)$, where
$f_I=f_{min}+IB$,
($I=0,\ldots,M-1$).  Such an operation moves the spectrum of the data towards
zero by frequency $f_I$.  We then apply a low pass filter with a cutoff
frequency $B$ and we resample the resulting sequence with the frequency $2B$.
The result is $M$ time series sampled at frequency $2 B$ instead of one sampled
at $2f_{\text{max}}$.  The resampled sequences are shorter than the original
ones by a factor of $M$ and each can be processed by a separate computer.  We
only need to perform the signal splitting operation once before the signal
search.  The splitting operation can also be performed continuously when the
data are collected so that there is no delay in the analysis.  The signal
splitting does not lead to a substantial saving in the total computational
power but yields shorter data sequences for the analysis.  For example for the case
of 7 days of observation time and sampling rate of 1 kHz the data itself would
occupy around 10 GB of memory (assuming double precision) which is available on
expensive supercomputers whereas if we split the data into a bandwidth of 50 Hz
so that sampling frequency is only 100 Hz each sequence will occupy 0.5 GB
memory which is available on inexpensive personal computers.
Analysis of these short sequences can be distributed over a large number
of small computers with no need of communication between the processors.
\subsection{Fine search}\label{subsec:fine}
To estimate the signal parameters we need to find
the global maximum of the statistics $\mathcal F$. We propose an
algorithm that consists of two parts: a {\em coarse} search and a
{\em fine} search. The coarse search involves calculation of $\mathcal F$
on an appropriate grid in parameter space and finding the maximum
value of $\mathcal F$ on the grid and the values of the parameters of the signal
that give the maximum. This gives coarse estimators of the
parameters. Fine search involves finding the maximum of $\mathcal F$
using fast maximization routines with the starting values
determined from the coarse estimates of the parameters. The grid
of the coarse search is determined by the region of convergence of
the maximization routine. We have tested extensively by means of
Monte Carlo simulation one such maximization algorithm - the
Nelder-Mead simplex method~\cite{P3}. This method applies to
function of arbitrary number of parameters and does not involve
calculation of the derivatives of the function. In Figure 1 we
illustrate our method for the case of a signal with just one
spindown parameter $f_1$ so that we need to find the maximum
of $\mathcal F$ with respect to frequency $f_0$ and $f_1$.
The use of the maximization procedure leads to an increase
in the significance of the detected events and an improved
accuracy of the estimation of the parameters of the signal.
It does not involve an increase in the complexity of the computation with
respect to a one-step search over a grid of templates. The
complexity is still of the order of $N_F (N \log_2N + L)$, where
$N_F$ is the number of templates in the grid, $N$ is the number of
points in the data sequence (we assume we use FFT to calculate
our statistics) and $L$ is some small number.

\section*{Acknowledgments}
I would like to thank Piotr Jaranowski for helpful discussions.
This work was supported by Polish Science Committee grant KBN 2 P303D 021 11.

\begin{figure}[ht]
\begin{picture}(10,13.5)
\put(0,0){\includegraphics{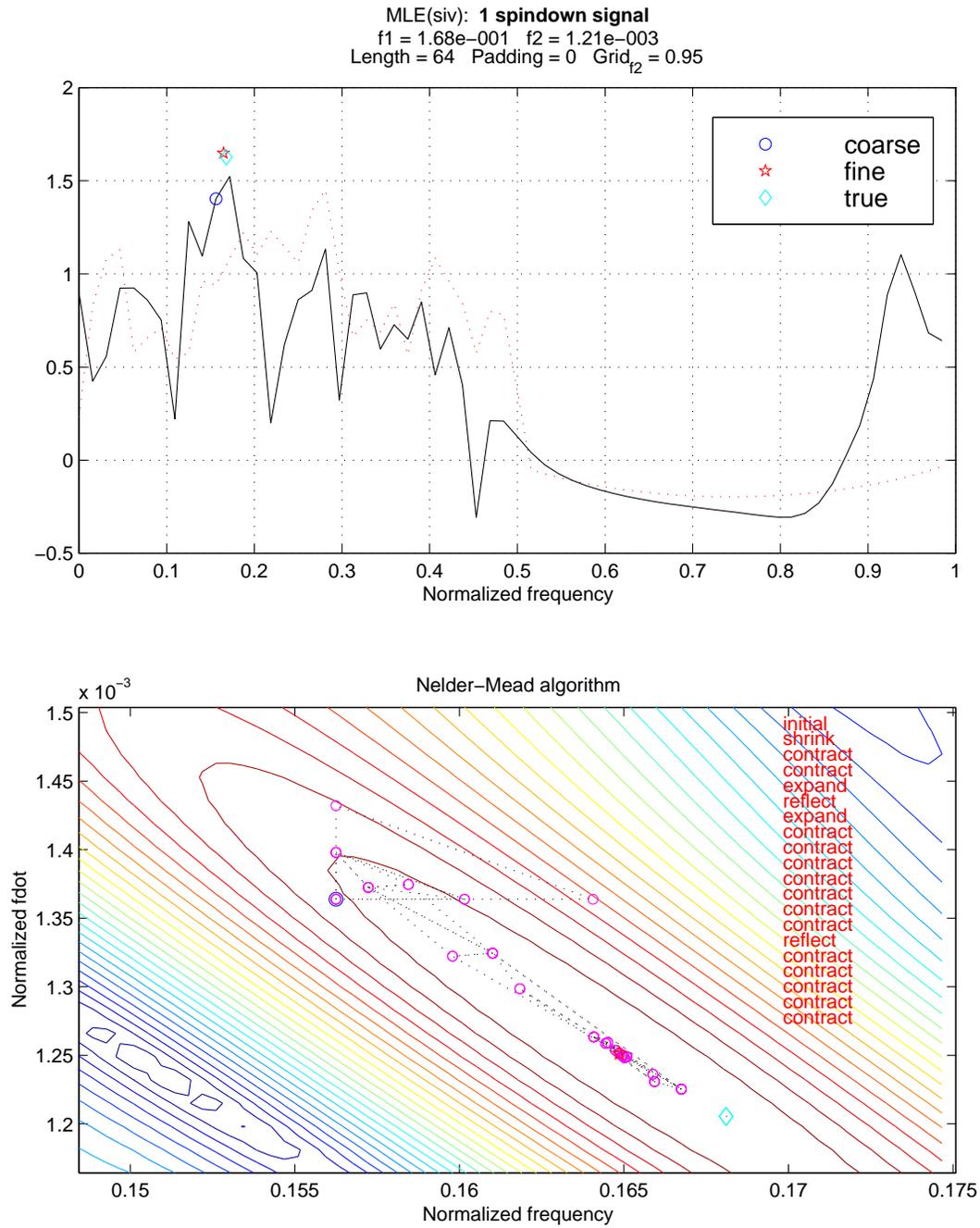}}
\end{picture}
\vspace{6.5in}
\caption{A two-step procedure to find the maximum of the
statistics ${\mathcal F}$ to estimate the parameters of the
signal. A search over a coarse grid gives a maximum marked by a
circle and a fine search using the Nelder-Mead algorithms gives
the maximum marked by a star. The true maximum is marked by a
diamond. The upper plot gives the magnitude of the FFT
for data before filtering (dotted line) and after filtering
(continuous line) implied in the definition of ${\mathcal F}$.
The lower plot is a contour plot of the statistics ${\mathcal F}$
as a function of frequency $f_0$ and spindown parameter $f_1$.
The steps of the simplex algorithm are displayed.
\label{fig:fNM}}
\end{figure}

\end{document}